\documentclass[prd,twocolumn,showpacs,floatfix,superscriptaddress]{revtex4}
\usepackage{graphicx}
\usepackage{amsmath}
\usepackage{amssymb}
\usepackage{bm}
\begin {document}

\title{Quark contribution to r-mode instabilities for several phases of deconfined quark matter}

\author{Basil A. Sa'd}
\email{sad@fias.uni-frankfurt.de}
\affiliation{%
Frankfurt Institute for Advanced Studies,
J.W. Goethe-Universit\"{a}t,
D-60438 Frankfurt am Main, Germany}%
\affiliation{Institut f\"ur Theoretische Physik,
J.W.\ Goethe-Universit\"at,
D-60438 Frankfurt am Main, Germany}




\date{\today}

\begin{abstract}
R-mode instabilities lead to specific signatures in the evolution of rotating pulsars, and may provide a unique signature which can identify the phase of matter that exists in the interiors of such objects.

 The contributions of quarks' bulk and shear viscosities to the dissipation of r-mode instabilities are studied. It is shown that (contrary to earlier works) the quark contribution is \emph{not} fully suppressed in the presence of fully gapped color-superconducting phase due to the fact that the bulk viscosity is a resonance effect between the density oscillation on one side, and the interaction rates that try to restore $\beta$-equilibrium during these oscillations on the other. This will have an effect on the structure of the r-mode instability window and will affect the angular momentum evolution of pulsars.
\end{abstract}



\maketitle

\section{Introduction}

 Compact (neutron) stars provide a natural laboratory of matter under 
extreme conditions. In the central regions of such stars the baryon 
density of matter could reach values up to 10 times the nuclear 
saturation density (i.e., $10 \rho_0$ where $\rho_0\simeq 
0.15~\mbox{fm}^{-3}$). At such high densities matter is likely to 
be in a deconfined state in which quarks, rather than hadrons, are the 
natural dynamical degrees of freedom \cite{Ivanenko1965,Ivanenko1969,
Itoh1970,Iachello1974,Collins1975}. At sufficiently low temperatures, the ground state of deconfined quark matter is a color 
superconductor. (For reviews on color superconductivity, see Refs.~\cite{RajWil, Alfordreview,Reddy2002,Rischke2003,Buballa2003,Huangreview,Shovkovy2004, Alford:2006fw}). Many phases of color superconductivity are known 
that could possibly be realized in dense matter. It remains unclear, however, 
which of these describe the ground state of matter under the specific conditions 
in stars. This is because of theoretical uncertainties in treating the 
strongly coupled, non-perturbative dynamics in QCD at the baryon densities of relevance.

 There are many instabilities associated with a star's rotation that can be operative in a neutron star. The most familiar of those is that if the star rotates with frequencies above a critical value, known as the Kepler frequency, mass shedding sets in from the star's equator. This frequency is $\nu_K \sim 1200 \mbox{Hz} $ in a typical neutron star. 

 Of concern to us are the r-mode instabilities. (For reviews on the r-mode instabilities see REfs.~\cite{Anderssonreview, Lind-lect}). R-modes are fluid waves in rotating stars with the Coriolis force acting as the restoring force. These waves couple to gravitational radiation emission causing the star to lose angular momentum. Only when dissipative phenomena dampen these r-modes can the star rotate without losing any angular momentum. This sets a limit on the maximum rotation frequency of the star at any given temperature. This also affects the rotation frequency evolution of the star, see, for instance Ref.~\cite{Zheng:2006gm}.
 
 There can be many dissipative phenomena that will damp the r-modes, these phenomena depend strongly on the phase of matter expected and on the structure of the star. In this paper we study the quark contribution to the r-mode instability window of quark stars for different phases of deconfined quark matter.

 At first, we consider the r-mode instability window for unpaired quark matter in comparison with the r-mode instability window of proton-neutron-electron matter (\emph{npe} matter), the results are very similar to the earlier work in Ref.~\cite{Madsenprl2}.

 It was suggested \cite{Madsenprl2} that the gapped quark contribution in the color-flavor locked (CFL) phase is strongly suppressed even if the energy gap is as small as $\Delta_{CFL} \sim 1 \mbox{MeV}$. This will be reconsidered more carefully in this article. Moreover, the r-mode instabilities for the two-flavor color superconductor (2SC) phase will be studied, the results are different from those of Ref.~\cite{Madsenprl2} due to the fact that the bulk viscosity may increase with the reduction of weak interaction rates as was seen in Ref.~\cite{SSR1}. The quark contribution to the r-mode instability window in the color-spin-locked (CSL) phase will also be studied.

 This paper is organized as follows: A quick review of the r-mode instabilities is presented in Sec. \ref{rm}. Quarks contribute to the r-mode instability window mainly through two dissipative phenomena, the bulk viscosity and the shear viscosity, these are reviewed in Sec. \ref{dp}. In order to relate the dissipative factors ``the bulk and shear viscosities" to the r-mode instabilities, one has to calculate the dissipation timescales related to these factors, these are discussed in Sec. \ref{dts}. The results for the r-mode instability window are shown in Sec.~\ref{results}. The discussion of the results is given in Sec.~\ref{d}. 

\section{R-modes}
\label{rm}
 Non-radial pulsations of the star couple to gravitational radiation (GR) emission and the GR emitted carries away energy and angular momentum from the star. In a non-rotating star, the effect of emitted GR is to dampen these oscillations. In a rotating star, however, the situation is rather different. The emission of GR causes the modes to grow, the reason being that modes which propagate in the direction opposite to the star's rotation have \emph{negative} angular momentum as seen in the co-rotating frame. Therefore, GR lowers the already negative angular momentum, \emph{i.e.}, the mode accelerates in the opposite direction of the star's rotation and the energy of the mode grows, \emph{i.e.}, the amplitude of the mode grows.

 R-modes are fluid waves on rotating stars with the Coriolis force acting as the restoring force. The r-modes are primarily velocity perturbations, which are solutions to the perturbed fluid equations. For a slowly rotating star they have the form \cite{Anderssonreview, Lind-lect, lindblom_owen_morsink,lindblom_andersson}

\begin{equation}
\delta \vec{v}= \alpha R \Omega \left( \frac{r}{R} \right) ^l \vec{Y}^{B}_{lm}e^{i \omega t},
\label{velocity_pert}
\end{equation} 

\noindent where $R$ and $\Omega$ are the radius and angular velocity of the unperturbed star, $\alpha$ is an arbitrary dimensionless amplitude of the mode, and $\vec{Y}^{B}_{lm}$ is the magnetic-type spherical harmonic defined by

\begin{equation}
\vec{Y}^{B}_{lm} \equiv \frac{r\vec{\nabla}\times (r\vec{\nabla} Y_{lm})}{\sqrt{l(l+1)}}.
\end{equation}

 Because the Coriolis force dominates, the frequencies of the r-modes are independent of the equation of state and are proportional to the angular velocity of the star \cite{Anderssonreview, Lind-lect, lindblom_owen_morsink,lindblom_andersson}
\begin{equation}
\omega =-\frac{(m-1)(m+2)}{m+1}\Omega.
\label{r-m-freq}
\end{equation}
 
\noindent The expressions for $\delta \vec{v}$ and $\omega$ are the lowest-order terms in $\Omega$. 
 
 Since r-modes are generally velocity perturbations, the energy of the mode has both kinetic and gravitational potential energy terms. We follow the Cowling approximation, in which the perturbation of the gravitational potential can be neglected. This has been shown to be a very good approximation, see, for instance, Refs. \cite{Anderssonreview, Lind-lect, lindblom_owen_morsink,lindblom_andersson}. The energy of the mode (measured in the co-rotating frame) is

\begin{equation}
\tilde{E}=\frac{1}{2} \int \rho (\delta\vec{v}^\ast \cdot \delta\vec{v})d^3x.
\label{r-m-energy}
\end{equation}

 This energy is conserved in the absence of dissipation. Including dissipation it satisfies \cite{Anderssonreview, Lind-lect, lindblom_owen_morsink}

 \begin{eqnarray}
  \frac{d\tilde{E}}{dt}=&-&\omega(\omega+m\Omega)\sum_{l\geq m}N_l \omega^{2l} \left[ \vert \delta D_{lm}\vert^2+\frac{4l\vert \delta J_{lm} \vert^2}{c^2(l+1)} \right] \nonumber \\
&-&\int \left[2\eta ( \delta \sigma_{ab}^{\ast} \delta \sigma^{ab})+\zeta ( \delta \sigma ^\ast \delta \sigma)\right]d^3x.
 \label{r-m-edot}
 \end{eqnarray}
 The second term in Eq.~(\ref{r-m-edot}) represents the dissipation due to viscosities of the fluid. It can be seen from the \emph{negative} sign in front of the term that the effect is always to reduce the energy of the mode converting it to heat. The thermodynamic functions $\eta$ and $\zeta$ are the shear and bulk viscosities of the fluid. The viscous forces are driven by the shear $\delta \sigma_{ab}$ and the expansion scalar $\delta \sigma$ of the perturbation defined by

\begin{eqnarray}
\delta \sigma_{ab}&\equiv&\frac{1}{2}(\nabla_a \delta v_b+\nabla_b \delta v_a-\frac{2}{3}\delta_{ab} \nabla_c \delta v^c),\\
\label{shear-pert}
\delta \sigma &\equiv& \nabla_a \delta v^a.
\label{bulk-pert}
\end{eqnarray}

 The first term in Eq.~(\ref{r-m-edot}) represents the effect of GR; $\delta D_{lm}$ and $\delta J_{lm}$ are the mass and current multipole moments of the perturbation,
\begin{eqnarray}
\delta D_{lm}=\int \delta \rho r^l Y_{lm}^\ast d^3x,\\
\label{mass_mult}
\delta J_{lm}=\int r^l (\rho \delta \vec{v}+\delta \rho \vec{v})\cdot \vec{Y}^{B\ast}_{lm} d^3x. 
\label{current_mult}
\end{eqnarray}
The constants $N_l=\frac{4 \pi G_N}{c^{2l+1}}\frac{(l+1)(l+2)}{(l(l-1))\lbrack (2l+1)!!\rbrack^2}$ are positive and the sum is also positive definite ($G_N=6.6742 \times 10^{-11} \mbox{m}^3 \mbox{Kg}^{-1} \mbox{s}^{-2}$ is the gravitational constant). Thus, the effect of GR is determined by the sign of $\omega (\omega + m\Omega)$, which is the product of the frequencies of r-modes in the inertial and the rotating frame,

\begin{equation}
\omega(\omega+m\Omega)=-\frac{2(m-1)(m+2)}{(m+1)^2}\Omega^2<0,
\end{equation}
 which implies that the total sign of the first term in Eq~(\ref{r-m-edot}) is always positive. GR always increases the energy of the mode.

 It has been shown \cite{Anderssonreview, lindblom_owen_morsink, lindblom_andersson, kokkotas_stergioulas} that only modes with $l=m$ can exist and that the most unstable mode is the one with $l=m=2$, which will be the one considered in our calculations.

\section{dissipative phenomena}
\label{dp}
 
 \subsection{Bulk viscosity}
 
 In general, the bulk viscosity is a measure of the kinetic energy dissipation 
during expansion and compression of a fluid. In compact stars, the
density oscillations of interest have characteristic frequencies that 
are of the same order of magnitude as the stellar rotation frequencies.
These are bound from below and from above, $1~\mbox{s}^{-1} 
\lesssim \omega\lesssim 
10^{3}~\mbox{s}^{-1}$. (For the fastest-spinning pulsar currently known, 
PSR J1748-2446ad, one has $\omega\approx 4.5\times 10^{3}~\mbox{s}^{-1}$ 
corresponding to $\nu=716~\mbox{Hz}$ \cite{716Hz}). Under conditions 
in stars, the bulk viscosity of quark matter is determined by the combined effect 
of the flavor-changing weak processes. When an (instantaneous) departure from 
chemical equilibrium is induced by expansion/compression of 
matter ($\delta \mu_1 = \mu_s-\mu_d$, and $\delta \mu_2 = \mu_s - \mu_u - \mu_e$), the weak processes try to restore the equilibrium state ($\delta \mu_i = 0$) and, 
while doing this, reduce the oscillation energy. 

\begin{figure}
\begin{center}
\noindent
\includegraphics[width=0.3\textwidth]{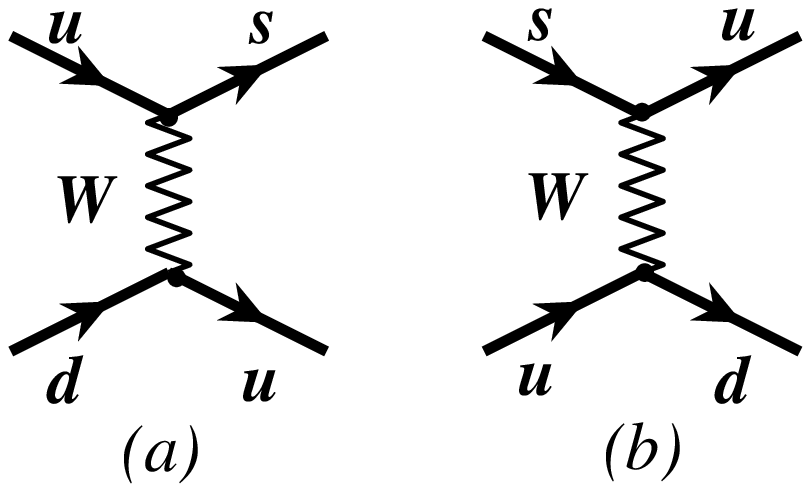}\\
\includegraphics[width=0.3\textwidth]{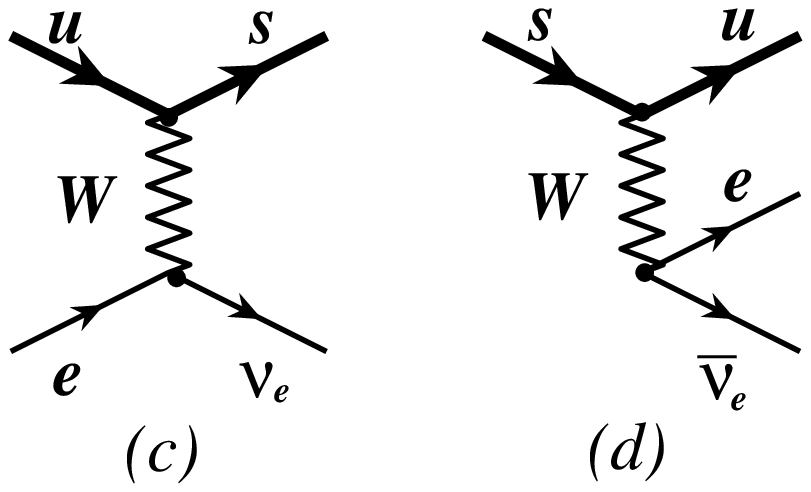}\\
\includegraphics[width=0.3\textwidth]{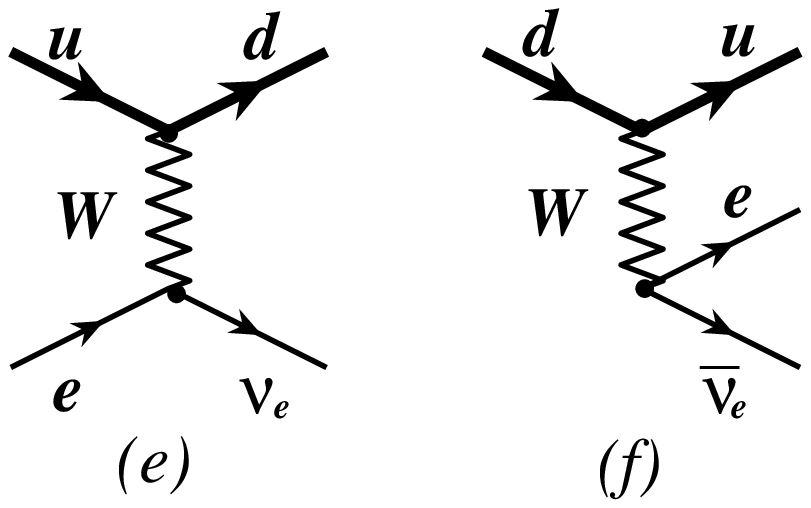}
\caption{Diagrammatic representation of the weak processes
that contribute to the bulk viscosity of quark matter in stellar cores.}
\label{fig-Urca_d_u_e}
\end{center}
\end{figure}

These weak interaction are: the non-leptonic weak 
processes $u + d \leftrightarrow u + s$ shown diagrammatically in Figs.~\ref{fig-Urca_d_u_e}(a) and 
\ref{fig-Urca_d_u_e}(b), and  the Urca (semi-leptonic) processes $d(s) \rightarrow u+e+\bar{\nu}_e,~ u+e\rightarrow d(s)+\nu_e$, see Figs.~\ref{fig-Urca_d_u_e}(c)--\ref{fig-Urca_d_u_e}(f). 

For studies of the viscosity in various phases of dense nuclear matter, see Refs.~\cite{FlowersItoh1,FlowersItoh2,Sawyer,Jones1,Lindblom1,Lindblom2,Drago1,Haensel1,Haensel2,Chat,Alford:2006gy,Sawyer2,Madsen,Wang1984,Xiaoping:2005js,Xiaoping:2004wc,Zheng:2002jq,Dai:1996fe, SSR1,SSR2}.
 
 The bulk viscosity due to weak interactions in quark matter is given by, see Ref.~\cite{SSR2} for a detailed derivation, 
\begin{equation}
\zeta=\zeta_{1}+\zeta_{2}+\zeta_{3},
\label{zeta-general}
\end{equation}

where

\begin{widetext}
\begin{subequations}
\begin{eqnarray}
\hspace{-20mm} \zeta_{1} &=& \frac{n}{\omega}\frac{\alpha_{2}\alpha_{3}}{g_{1}^2+g_{2}^2}
\left[\alpha_{1}\alpha_{2}\alpha_{3} C_{1}^2
+\left(\alpha_{1}+\alpha_{2}+\alpha_{3}\right)
\left(A_{1}C_{2}-A_{2}C_{1}\right)^2\right],
\label{zeta11} \\
\hspace{-20mm} \zeta_{2} &=& \frac{n}{\omega}\frac{\alpha_{1}\alpha_{3}}{g_{1}^2+g_{2}^2}
\left[
\alpha_{1}\alpha_{2}\alpha_{3} C_{2}^2
+\left(\alpha_{1}+\alpha_{2}+\alpha_{3}\right)
\left[\left(A_{2}-B_{2} \right) C_{1} -A_{2}C_{2}\right]^2\right],
\label{zeta22} \\
\hspace{-20mm} \zeta_{3} &=& \frac{n}{\omega}\frac{\alpha_{1}\alpha_{2}}{g_{1}^2+g_{2}^2}
\left[
\alpha_{1}\alpha_{2}\alpha_{3} \left(C_{1}-C_{2}\right)^2
+\left(\alpha_{1}+\alpha_{2}+\alpha_{3}\right)
\left(B_{1}C_{2}-B_{2}C_{1}\right)^2\right].
\label{zeta33}
\end{eqnarray}
\label{zeta1-3}
\end{subequations}
\end{widetext}

\noindent with $\alpha_i \equiv n \omega / \lambda_i$, $n$ is the baryon density, $\omega$ is the oscillation frequency, and $\lambda_i$
 is the difference of the forward and reveres weak interaction rates to first order in $\delta \mu_i$.

 The thermodynamic functions $A_i$, $B_i$, and $C_i$ are function of the chemical potentials and their derivatives with respect to the baryon number density, they are discussed in detail in Ref.~\cite{SSR2}, they do not vary significantly whether in the normal phase or in a color-superconducting phase as was seen in Ref.~\cite{SSR1}.

In order to further proceed with the calculation of the bulk viscosity 
in the normal phase of three-flavor quark matter, we also need to know 
the rate difference of the three pairs of weak processes in 
Fig.~\ref{fig-Urca_d_u_e}.

\begin{subequations}
\begin{eqnarray}
\Gamma_{(a)} - \Gamma_{(b)} &=& - \lambda_{1} \delta\mu_{1} ,
\label{ratediff1}\\
\Gamma_{(c)} - \Gamma_{(d)} &=& - \lambda_{2} \delta\mu_{2} ,
\label{ratediff2}\\
\Gamma_{(e)} - \Gamma_{(f)} &=& - \lambda_{3} \left(\delta\mu_{2}-\delta\mu_{1}\right) .
\label{ratediff3}
\end{eqnarray}
\label{ratediff0}
\end{subequations}

 The rates of both the non-leptonic and 
semi-leptonic processes have been calculated in the literature, 
see Refs.~\cite{Madsenrate,Sawyer2,Wang1984,Anand} and 
Refs.~\cite{Iwamoto,Iwamoto2}, respectively. In the limit of three 
massless quarks, for example, the rates are
\begin{subequations}
\begin{eqnarray}
\lambda_{1} &\simeq& \frac{64}{5\pi^3} G_F^2
                     \cos^2 \theta_C \sin^2 \theta_C \mu_d^5 T^2 ,
\label{lambda1}\\
\lambda_{2} &\simeq& \frac{17}{40\pi} G_F^2
                     \sin^2 \theta_C \mu_s m_s^2 T^4 ,                 
\label{lambda2}\\
\lambda_{3} &\simeq& \frac{17}{15\pi^2} G_F^2
                     \cos^2 \theta_C \alpha _s \mu_d \mu_u \mu_e T^4 .
\label{lambda3}
\end{eqnarray}
\label{lambda123}
\end{subequations} 

 Before we proceed, it is instructive to discuss the effect of color superconductivity on $\lambda_i$'s. Recall that $\Gamma_i(\delta \mu_i,~\Delta)-\bar{\Gamma}_i(\delta \mu_i,~\Delta)=\lambda_i(\Delta)\delta \mu_i$, see Eqs.~(\ref{ratediff0}) and that $\Gamma_i(\delta \mu_i,~0)=\bar{\Gamma}_i(-\delta \mu_i,~0)$ where $\Gamma_i(\delta \mu_i, \Delta)$ is the forward interaction rate as a function of both $\delta \mu_i$ and $\Delta$. $\bar{\Gamma}_i(\delta \mu_i,~\Delta)$ is the inverse interaction rate (backwards channel). For instance, if we choose $\Gamma_{(a)}$ in Fig.~\ref{fig-Urca_d_u_e} and Eqs.~(\ref{ratediff0}) as the forward channel $\Gamma_1(\delta \mu_1,~\Delta)$, $\bar{\Gamma}_{1}$ would be  $\Gamma_{(b)}$ in Fig.~\ref{fig-Urca_d_u_e} and so forth. From the two equations mentioned above it is easy to show that $\Gamma_i(\delta \mu_i,0)=-\bar{\Gamma_i}(\delta \mu_i, 0)=\frac{1}{2}\lambda_i(0)\delta \mu_i$.

 Color superconductivity introduces a reduction factor $H_i(\Delta)$ to the interaction rate $\Gamma_i$ and another suppression factor $\bar{H}_i(\Delta)$ to the inverse interaction,

\begin{eqnarray}
\Gamma_i(\delta \mu_i, \Delta)&=&\Gamma_i(\delta \mu_i, 0)H_i(\Delta),\\
\bar{\Gamma_i}(\delta \mu_i, \Delta)&=&\bar{\Gamma_i}(\delta \mu_i, 0)\bar{H_i}(\Delta).
\end{eqnarray}

 The relation between the gapped $\lambda_i(\Delta)$ and  the ungapped $\lambda_i(0)$ becomes,

\begin{eqnarray}
\lambda_i(\Delta)\delta \mu &=& \Gamma_i(\delta \mu_i,~0)H_i(\Delta)-\bar{\Gamma}_i(\delta \mu_i,~0)\bar{H}_i(\Delta)\nonumber \\
&=& \Gamma_i(\delta \mu_i,~0)H_i(\Delta)-\Gamma_i(-\delta \mu_i,~0)\bar{H}_i(\Delta)\nonumber \\
&=& \lambda_i(0)\frac{H_i(\Delta)+\bar{H}_i(\Delta)}{2}\delta \mu.
\end{eqnarray}

 Then the reduction to $\lambda_i(0)$ would be a function of the suppressions to both the forward and the inverse interactions.
\begin{equation}
\lambda_i(\Delta)=\lambda_i(0)\frac{H_i(\Delta)+\bar{H}_i(\Delta)}{2}.
\label{reduce}
\end{equation}

 In the following, we will approximate the reduction $H(\Delta)$ to be of an exponential form for each incoming gapped quark branch in the interaction.

\subsection{Shear Viscosity}

 The shear viscosity of ungapped quark matter is dominated by quark-quark scattering, for ungapped quark matter it is given by \cite{GondekRosinska:2003iy, Heiselberg:1993cr}
\begin{equation}
\eta = 5 \times 10^{15} \left(\frac{0.1}{\alpha_s}\right) \left(\frac{n}{\rho_0}\right)^{14/9} \left(\frac{T}{10^9 K}\right)^{-5/3} ~~{g cm^{-1} s^{-1}}.
\label{sqm-sv}
\end{equation}

 The effect of color superconductivity is to suppress the quark excitations exponentially, if all quark modes are gapped, we can neglect the quark shear viscosity from consideration since the shear viscosity will be reduced by approximately (including screening) $e^{-\frac{\Delta}{3 T}}$. In this case, the quark contribution to the r-mode instabilities is solely governed by the bulk viscosity. Otherwise, only ugapped quarks' contribution will be considered.

 For the remainder of this article the values $\alpha_s = 0.1$, $n=5\rho_0$ will be used.

\section{Dissipative Timescales}
\label{dts}
 In order to compare the relative strengths of dissipative phenomena and GR emission it is convenient to calculate the dissipative timescales associated with these effects defined as 

\begin{equation}
\tau_i \equiv -\frac{2 \tilde{E}}{(d\tilde{E}/dt)_i},
\end{equation}
where the index $i$ can stand for any dissipative phenomenon such as bulk viscosity (bv), shear viscosity (sv), or GR emission (GR).

 The energy of the mode can be found using Eqs.~(\ref{velocity_pert}) and (\ref{r-m-energy}). Assuming spherical symmetry \cite{Anderssonreview, Lind-lect, lindblom_owen_morsink},
 
\begin{equation}
 \tilde{E}=\frac{1}{2}\alpha^2 \Omega^2 R^{-2l+2}\int^R \rho r^{2l+2} dr.
 \label{energy-of-mode}
\end{equation}

\subsection{Gravitational Radiation Timescale}
 
 The lowest-order contribution to the GR terms in Eq.~(\ref{r-m-edot}) comes entirely from the current multipole moment $\delta J_{ll}$. This can be evaluated using Eqs.~(\ref{velocity_pert}) and (\ref{current_mult}), to lowest order in $\Omega$ \cite{Anderssonreview, Lind-lect, lindblom_owen_morsink}

\begin{equation}
\delta J_{ll}=\frac{2 \alpha \Omega}{c R^{l-1}}\sqrt{\frac{l}{l+1}}\int^R \rho r^{2l+2} dr.
\end{equation}

 This leads to a dissipative timescale \cite{lindblom_owen_morsink, lindblom_andersson, Anderssonreview, kokkotas_stergioulas}
\begin{equation}
 \frac{1}{\tau_{GR}}=-\frac{32 \pi G \Omega^{2l+2}}{c^{2l+3}}\frac{(l-1)^2l}{\lbrack(2l+1)!!\rbrack^2} \left( \frac{l+2}{l+1} \right) ^{2l+2} \int^R \rho r^{2l+2} dr.
\end{equation}

\noindent Note that we are only interested in the case where $l=m=2$. Assuming a uniform density star the dissipative ``growth'' time for GR emission becomes
\begin{equation}
\frac{1}{\tau_{GR}}=-\frac{1}{3.26} \left( \frac{\Omega^2}{\pi G \rho} \right) ^2.
\label{taugr}
\end{equation}

\subsection{Shear Viscosity Timescale}
The time derivative of the mode energy due to the shear viscosity coefficient can be calculated from  Eqs.~(\ref{velocity_pert}), (\ref{r-m-edot}), and (\ref{shear-pert}), which leads to \cite{Anderssonreview,lindblom_owen_morsink, lindblom_andersson,kokkotas_stergioulas}

 \begin{equation}
 \frac{1}{\tau_{sv}}=(l-1)(2l+1)\frac{\int^R \eta r^{2l}dr}{\int^R \rho r^{2l+2}dr}.
 \end{equation}

\noindent Using $l=2$ and constant density, we get,

\begin{equation}
\frac{1}{\tau_{sv}}= \frac{28 \pi }{3}\frac{\eta R}{M},
\label{tau-sv}
\end{equation}
where $M$ is the mass of the star.

\subsection{Bulk Viscosity Timescale}

The time derivative of the co-rotating frame energy $\tilde{E}$ due to the effect of bulk viscosity is

\begin{equation}
\left(\frac{d\tilde{E}}{dt}\right)_{bv}=-\int \zeta \vert \vec{\nabla} \cdot \delta \vec{v} \vert^2 d^3x.
\label{1}
\end{equation}

 The reduction of $d\tilde{E}/dt$ to a one-dimensional integral is not straight forward; the expansion scalar of the mode $\vec{\nabla} \cdot \delta \vec{v}$ is a complicated function of radius and angle. However, since the bulk viscosity generally has no angular dependence, we may convert Eq.~(\ref{1}) to a one-dimensional integral by defining an angle averaged expansion scalar squared $\left< \vert \vec{\nabla} \cdot \delta \vec{v} \vert ^2 \right>$,

\begin{equation}
\left(\frac{d\tilde{E}}{dt}\right)_{bv}=- 4 \pi \int_0^R\zeta \left< \vert \vec{\nabla} \cdot \delta \vec{v} \vert^2 \right> r^2 dr.
\label{2}
\end{equation}

 The angle-averaged scalar expansion is in general a complicated function, it has only been determined numerically in Ref.~\cite{Lind2nd}, however, the simple analytical expression

\begin{equation}
\left< \vert \vec{\nabla} \cdot \delta \vec{v} \vert ^2 \right>=\frac{\alpha^2 \Omega^2}{690} \left( \frac{r}{R} \right)^6 \left[ 1+0.86 \left( \frac{r}{R} \right)^2 \right] \left( \frac{\Omega^2}{\pi G_N \rho} \right)^2
\label{3}
\end{equation}

 is an excellent fit to those numerical solutions \cite{Lindblom1}.

 Using Eqs.~(\ref{energy-of-mode}), (\ref{2}), and (\ref{3}) we get

\begin{equation}
\frac{1}{\tau_{bv}} \equiv -\frac{(d \tilde{E}/dt)_{bv}}{ \tilde{E}} \simeq 4.828 \times 10^{-2} \left( \frac{\Omega ^2}{\pi G_N \rho} \right)^2 \frac{\zeta}{\rho R^2}.
\label{tau-bv}
\end{equation}

 The critical angular velocity above which GR has the shortest timescale is found by solving \cite{Anderssonreview, Lind-lect, lindblom_owen_morsink}

\begin{equation}
\frac{1}{\tilde{E}}\frac{d \tilde{E}}{dt}=\frac{1}{\tau_{GR}}+\sum \frac{1}{\tau_{diss}}=0.
\label{r-m-instability}
\end{equation}
 
\noindent Where $\frac{1}{\tau_{diss}}=\Sigma \frac{1}{\tau_{V}}$ is the sum of the inverse of the dissipative timescales (but not the gravitational radiation timescale). Above the critical frequency the growth of the r-modes due to GR emission is the most dominant effect and the star emits GR and loses angular momentum  until the effect of viscosities is the dominant one.

\section{The R-mode Instability Window for Quark Stars}
\label{results}

 The dominant dissipative phenomena in a quark star are the quark bulk and shear viscosities. There are other dissipative phenomena that can play a role in determining the r-mode instability window, for example, the shear viscosity due to electron-electron scattering, or, in the case of a superfluid phase, the friction between the normal and the superfluid component that adds other viscosity coefficients \cite{landau_fluid}. We are concerned with the contribution of quarks to the r-mode instability window.

Our results are also of relevance to hybrid stars. However, to get a complete picture, the effect of hadrons need to be considered as well, and friction between the quark core and the hadronic mantle will also play a decisive role in finding the r-mode instability window.

\subsection{R-mode Instability Window for the Normal Phase}
 
\begin{figure}
\noindent
\includegraphics[width=0.9\linewidth]{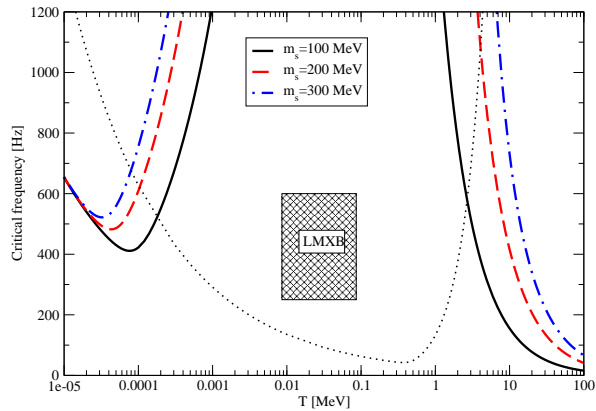}\\~\\~\\
\caption{(color online) The r-mode instability window for normal quark stars (thick lines) for strange quark masses of $100$, $200$, and $300$ MeV, and neutron stars (thin dotted line) \cite{Lind-lect}. The shaded box represents the region in which most LMXB's are observed.}
\label{nu_c_normal}
\end{figure}

 Using Eqs.~(\ref{zeta1-3}) and (\ref{sqm-sv}) alongside with Eqs.~(\ref{taugr}), (\ref{tau-sv}), and (\ref{tau-bv}) one can calculate the R-mode instability window from Eq.~(\ref{r-m-instability}). The results for the critical frequencies are shown in Fig.~\ref{nu_c_normal}, the region above the curve is the unstable region towards the emission of GR from r-modes. For comparison, the r-mode instability window for neutron-proton-electron matter (\emph{npe} matter) \cite{Lind-lect} is shown.  

 At low temperatures $T<0.1 \mbox{keV}$ the shear viscosity of quark matter starts to be significant. Above that temperature the bulk viscosity is responsible for the stability of the star. At temperatures above $1-10 \mbox{MeV}$, where the bulk viscosity starts dropping, the region is unstable towards GR emission from r-modes. 
 
 The observed distribution of Low Mass X-ray Binaries (LMXB's), presumed to be old pulsars spun up by accretion of matter from the binary companion to become rapid millisecond pulsars, is indicated in the plot by a shaded box. It can easily be seen that ordinary neutron star models place these pulsars within an unstable region, leading us to exclude the \emph{classical} model of neutron stars. On the other hand, the quark matter model places them within a stable region. One \emph{cannot}, however, exclude the possibility of \emph{npe} matter in these stars if one considers a more sophisticated neutron star model with a liquid core surrounded by a nearly static crust. This configuration leads to damping due to viscosity in the boundary layer between the oscillating fluid and the crust, which is about $10^5$ times bigger than the damping from the shear viscosity in the interior. This would place the LMXB's in a stable region, see, for instance, Ref. \cite{Madsenprl2}.
 
 \subsection{R-mode Instability Window for the CFL Phase}

 Color superconductivity changes the situation from that of unpaired quark matter. The severity of this change depends on the type of pairing and the value of the gap. The CFL phase gaps all quark modes. Here we consider the contribution of quarks on the bulk viscosity and r-mode instabilities of the CFL phase.

 The critical temperature might be as large as $T_c \simeq 50$ MeV when the quark chemical potential $\mu_s \sim 400$ MeV \cite{rapp_schaefer} leading to $\Delta_0^{CFL} \simeq 89$ MeV. The energy gap has a dependence on the temperature as $\Delta_{CFL}=\Delta_{0}^{CFL}\sqrt{1-\left( \frac{T}{T_c}\right)^2}$.

 We neglect the effect of color superconductivity on the thermodynamic coefficients $A_i$, $B_i$, and $C_i$. Therefore, the effects of the CFL phase are simply to reduce the $\lambda_i$'s as in Eq.~(\ref{reduce}). Non-leptonic weak interactions have two quarks on both the forward and backward channels, each of which will be suppressed exponentially, leading to $H_1(\Delta_{CFL})=\bar{H}_1(\Delta_{CFL}) \simeq e^{-\frac{2\Delta_{CFL}}{T}}$. Thus,

\begin{equation}
\lambda_1(\Delta_{CFL}) / \lambda_1(0) = e^{-\frac{2\Delta_{CFL}}{T}}.
\end{equation}

\noindent The two other neutrino-emitting interactions both have one quark flavor in both channels. Hence, the reduction to the interaction rates will take the form $H_{2,~3}(\Delta_{CFL})=\bar{H}_{2,~3}(\Delta_{CFL}) \simeq e^{-\frac{\Delta_{CFL}}{T}}$. Then the reduction of $\lambda_{2,~3}$ is

\begin{equation}
\lambda_{2,3}(\Delta_{CFL}) / \lambda_{2,3}(0) = e^{-\frac{\Delta_{CFL}}{T}}.
\end{equation} 
 
\begin{figure}
\begin{center}
\includegraphics[width=0.9\linewidth]{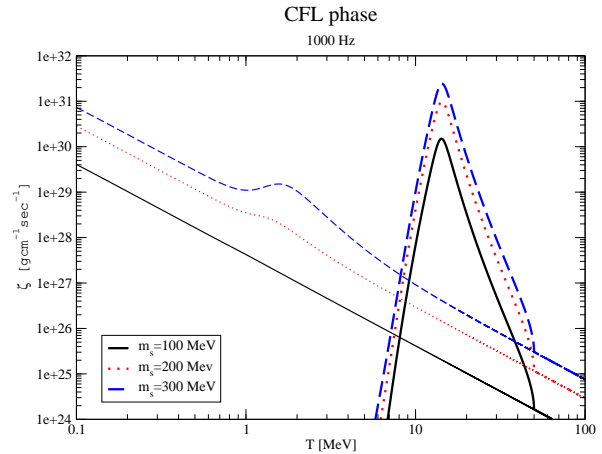}\\~\\~\\~\\
\includegraphics[width=0.9\linewidth]{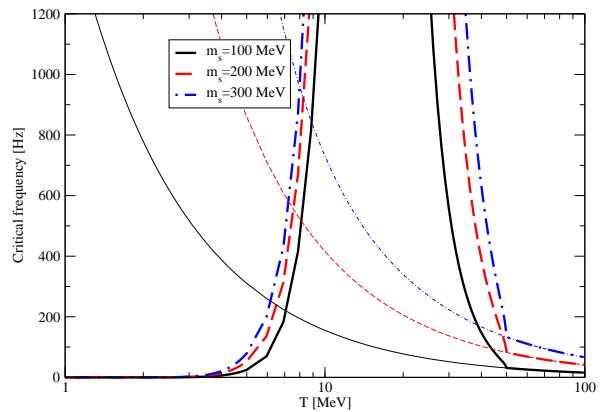}\\~\\
\caption{(color online) The quark contribution to the bulk viscosity as a function of temperature at constant frequency ``1000 Hz'' and r-mode instabilities of the CFL phase for different values of the strange quark mass. The thin lines show the results for unpaired quark matter.}
\label{fig:BV_nu_c_cfl}
\end{center}
\end{figure}

 We follow the same procedure as for normal quark matter to calculate the bulk viscosity and r-mode instability window. Since the shear viscosity only participates in the r-mode instability at low temperature and all quark flavors are strongly suppressed, we can safely neglect the quark shear viscosity. The results for a critical temperature $T_c=50 ~\mbox{MeV}$ are shown in Fig.~\ref{fig:BV_nu_c_cfl}

  Quarks allow for a narrow stability window at high temperatures $T>5$ MeV. This is due to the fact that color superconductivity may actually increase the bulk viscosity as can be seen in Fig.~\ref{fig:BV_nu_c_cfl}. 

 It is clear from Fig.~\ref{fig:BV_nu_c_cfl} that considering only the quarks' contribution to the stability window places all LMXB in an unstable region. However, one cannot yet draw any significant astrophysical conclusions; there are other dissipative phenomena besides just the bulk and shear viscosities due to quarks in the CFL phase, such as the bulk and shear viscosities due to superfluid phonons \cite{cristina_bulk, cristina_shear}, and the bulk viscosity due to kaons \cite{mark_bulk}. Without a complete analysis of all these phenomena, no significant conclusions can be drawn. However, we do notice that at large temperatures there is a narrow stability region. Incidentally, this is the region where young pulsars are born, which means that young quark stars in the CFL phase are produced in a stable region and then have to cool down before entering the instability region and slowing down.

 The result for the r-mode instability window differs significantly from that in Ref.~\cite{Madsenprl2}, which arises from the fact that the bulk viscosity may be enhanced rather than suppressed due to the suppression of the interaction rates. This enhancement was not considered in Ref.~\cite{Madsenprl2}.

\subsection{R-mode Instability Window for the 2SC Phase}
\label{sub:r-m-2sc}

 The strange quark mass $m_s\sim 100$ MeV is larger than the up and down quark masses $m_{u,~d}\lesssim 10$ MeV. Therefore, there will be a Fermi surface mismatch between the strange quark and the up and down quarks. In this case the up and down quarks can pair and form a condensate, the so-called 2-flavor color-superconducting (2SC) phase.
 
 In the 2SC phase, the red and green quarks acquire a gap $\Delta_{2SC}$, whereas there is no gap for the blue quark (the choice of color is arbitrary). The critical temperature is $T_c \sim 30$ MeV, leading to $\Delta_0^{2SC} \simeq 52.6$ MeV \cite{Schmitt:2002sc}.

 Since no global symmetries are broken, the 2SC phase is not a superfluid. And since there are ungapped quarks, they will dominate all interaction rates and the physical properties of the system. They will only be different from normal quark matter by a factor as will be shown below, see also Refs.~\cite{Alford:2006gy, Madsenprl2}.

 The effect of gap the 2SC phase on the bulk viscosity is less severe than in the CFL phase. The reason is that there is an ungapped color for the up and down quarks and that the strange quarks remain unpaired.

\begin{figure}
\begin{center}
\includegraphics[width=0.75\linewidth]{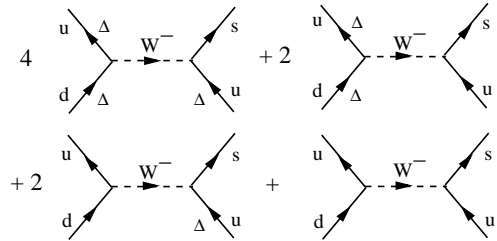}
\caption{Contributions to the process $u+s \leftrightarrow u+d$ in the 2SC phase. A gapped fermion is marked
with the gap $\Delta_{2SC}$ at the respective line \cite{Alford:2006gy}. }
\label{non_lept_decompose_2sc}
\end{center}
\end{figure}

 Since weak interactions do not change the quark colors at the vertices, we can decompose the non-leptonic weak interaction (the first interaction in Fig~\ref{fig-Urca_d_u_e}) into four different interactions according to the quark colors conserving the color at each vertex. This translates to a decomposition according to the gapped and ungapped modes of the quarks playing a role in the interaction as in Ref.~\cite{Alford:2006gy}, see Fig.~{\ref{non_lept_decompose_2sc}}.

 This leads to a reduction of the interaction rates $\Gamma_1$ and $\bar{\Gamma}_1$,

\begin{eqnarray}
H_1(\Delta_{2SC})&=&\frac{4}{9}e^{-\frac{2\Delta_{2SC}}{T}}+\frac{4}{9}e^{-\frac{\Delta_{2SC}}{T}}+\frac{1}{9},\\
\bar{H}_1(\Delta_{2SC})&=&\frac{6}{9}e^{-\frac{\Delta_{2SC}}{T}}+\frac{3}{9},
\end{eqnarray}

\noindent which gives

\begin{equation}
\lambda_1(\Delta_{2SC}) / \lambda_1(0) = \frac{2}{9}e^{-\frac{2\Delta_{2SC}}{T}}+\frac{5}{9}e^{-\frac{\Delta_{2SC}}{T}}+\frac{2}{9} .
\end{equation}
 
\noindent This gives a different fraction for the ungapped mode compared to that in Ref. \cite{Alford:2006gy}. This is due to their assumption that $\delta \mu_1>0$ which translated into saying that $\lambda_1(\Delta_2SC)=\lambda_1(0) H_1(\Delta_2SC)$. Assuming that $\delta \mu_i$ is positive neglects the difference between the suppressions of both directions of the interaction. 

 The same thing can be said about the other two weak interactions. However, since there is only one quark participating in the interaction in each channel, we decompose each interaction into two. the terms $\lambda_{2,~3}$ then take the form

\begin{eqnarray}
\lambda_2(\Delta_{2SC}) / \lambda_2(0) &=&  \frac{2}{3}+\frac{1}{3}e^{-\frac{\Delta_{2SC}}{T}}, \\
\lambda_3(\Delta_{2SC}) / \lambda_3(0) &=&  \frac{1}{3}+\frac{2}{3}e^{-\frac{\Delta_{2SC}}{T}} .
\end{eqnarray} 

 The gapped quarks' contribution to the shear viscosity can be neglected, then the reduction to the shear viscosity is only $(5/9)^{1/3}$ as in Ref~\cite{Madsenprl2}.

\begin{figure}
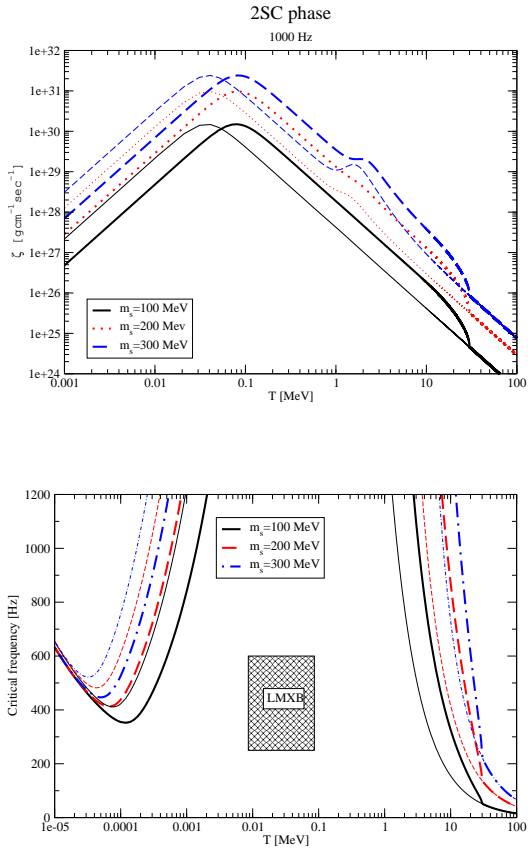

\begin{center}
\includegraphics[width=0.8\linewidth]{BV_2sc.eps}\\~\\~\\~\\
\includegraphics[width=0.8\linewidth]{nu_c_2sc.eps}\\~\\
\caption{(Color online) Same as Fig.~\ref{fig:BV_nu_c_cfl}, but for the 2SC phase.}
\label{fig:BV_nu_c_2sc}
\end{center}
\end{figure}

 Choosing the critical temperature to be $T_c=30$ MeV, the bulk viscosity and r-mode instability window can both be calculated. The results are shown in Fig.~\ref{fig:BV_nu_c_2sc}.

The results for the r-modes are also different from those in Ref.~\cite{Madsenprl2}. This is due to the assumption that the ungapped mode leads to a $1/9$ reduction of the interaction rate whereas we have shown that it should be $2/9$, and due to the fact that the reduction of the interaction rate has an effect on the bulk viscosity that is not as trivial as only suppressing it.

 It can be seen from Fig.~\ref{fig:BV_nu_c_2sc} that the effect of the 2SC phase is mainly to shift the bulk viscosity and r-mode instability window sideways by a multiplicative factor. LMXB data do not exclude the possibility of having a 2SC phase in pulsars.

\subsection{R-mode Instability Window for the CSL Phase}

 If the strange quark mass is large enough, the condition of charge neutrality requires that the electron chemical potential to be large. The strange quark mass and the large electron chemical potential will cause a substantial difference in particle species' Fermi momenta and pairing between different quark flavors may not be possible. The remaining option, then, is Cooper pairing of each flavor with itself.

 To maintain the fermionic antisymmetry of the Cooper pair wave function, single-flavor pairing phases have to be either symmetric in color, which greatly weakens or even eliminates attractive interactions, or symmetric in spin, which causes particles of the same spin to pair together and, therefore, the Cooper pairs will have a total spin of 1.

 For spin-1 color-superconducting phases which are discussed in Refs.~\cite{Schmitt:2004hg, SchmittSpin1, SSW2}, the situation is very similar to the 2SC phase since there are ungapped modes. However, it has been suggested \cite{ABBY} that for a particular spin-1 phase, the CSL phase, all modes are gapped. There are two modes gapped with an energy gap $\Delta_{CSL}$ and the ``ungapped'' mode is now gapped with $X_i\Delta_{CSL} \equiv \frac{m_i}{\sqrt{2}\mu_i} \Delta_{CSL}$ where $m_i$ and $\mu_i$ are the mass, and chemical potential of the quark flavor.

 Following the same procedure as for the 2SC phase, the weak interaction rates $\lambda_i$ take the form

\begin{widetext}
\begin{eqnarray}
\lambda_{1}(\Delta_{CSL}) / \lambda_{1}(0) &=& \frac{4}{9}e^{-\frac{2\Delta_{CSL}}{T}}+\frac{2}{9}e^{-\frac{(1+X_u)\Delta_{CSL}}{T}}+\frac{1}{9} \left( e^{-\frac{(1+X_d)\Delta_{CSL}}{T}}+e^{-\frac{(1+X_s)\Delta_{CSL}}{T}}\right)\nonumber\\
&+&\frac{1}{18} \left( e^{-\frac{(X_u+X_s)\Delta_{CSL}}{T}}+e^{-\frac{(X_u+X_d)\Delta_{CSL}}{T}} \right) ,\\
\lambda_{2}(\Delta_{CSL}) / \lambda_{2}(0) &=& \frac{1}{6} \left( e^{-\frac{X_s \Delta_{CSL}}{T}}+e^{-\frac{X_u \Delta_{CSL}}{T}}\right)+\frac{2}{3}e^{-\frac{\Delta_{CSL}}{T}},\\
\lambda_{3}(\Delta_{CSL}) / \lambda_{3}(0) &=& \frac{1}{6}\left(e^{-\frac{X_d \Delta_{CSL}}{T}}+e^{-\frac{X_u \Delta_{CSL}}{T}}\right)+\frac{2}{3}e^{-\frac{\Delta_{CSL}}{T}}.
\end{eqnarray}
\end{widetext}

\begin{figure}
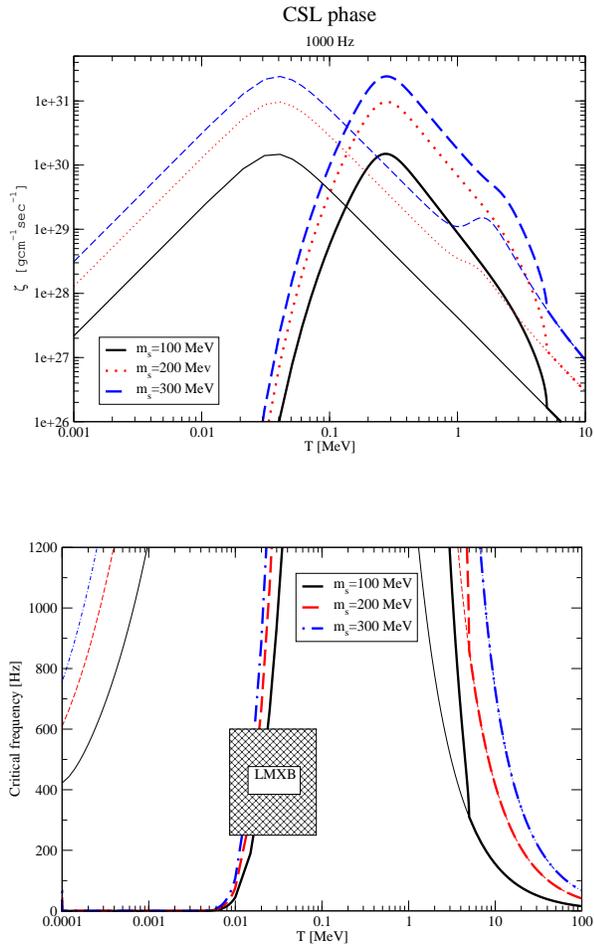

\begin{center}
\includegraphics[width=0.9\linewidth]{BV_csl.eps}\\~\\~\\~\\
\includegraphics[width=0.9\linewidth]{nu_c_csl.eps}\\~\\
\caption{(Color online) Same as Fig.~\ref{fig:BV_nu_c_cfl}, but for the CSL phase.}
\label{fig:BV_nu_c_csl}
\end{center}
\end{figure}

\noindent Choosing a critical temperature of $5$ MeV and $\frac{T_c}{\Delta_0}=0.8$ \cite{SchmittSpin1}, we calculate the bulk viscosity and the critical frequencies of the CSL phases. The results are shown in Fig.~\ref{fig:BV_nu_c_csl}. The effect of the quark shear viscosity has again been neglected since it only acts at low temperature and is then strongly suppressed.

 In Fig.~\ref{fig:BV_nu_c_csl} we see that the quark contribution places LMXB's partially within the window of stability. The situation will change, however, if we consider a smaller critical temperature than $5 \mbox{MeV}$. At a small enough $T_c$ LMXB's would be completely placed in a stable region.

 It must be noted that the CSL phase is a superfluid, and so, much like in the CFL phase, there are many other dissipative phenomena that need to be considered, such as the bulk and shear viscosities due to superfluid phonons, bulk viscosity due to Goldstone bosons, and friction between both superfluid and normal fluid components \cite{Gusakov:2007px}.

\section{discussion}
\label{d}
 In this article, the contribution of quarks to the r-mode instabilities was studied. The main conclusion is that one \emph{cannot} neglect the quark contribution to r-mode instabilities even if all quarks were gapped.

 It is shown in Fig.~(\ref{nu_c_normal}) that the r-mode instability window is very different for quark matter than for neutron star matter. This might provide us with an opportunity to distinguish between quark stars and neutron stars. Further investigation is clearly needed.

 Each different color-superconducting phase has a different structure for the r-mode instability window. Quarks in the CFL phase generate a narrow window of stability at temperatures which are, incidentally, the temperatures at which a young pulsar is formed, see Fig.~(\ref{fig:BV_nu_c_cfl}). This implies that a newly born quark star in the CFL phase is generated in a stable region, and has to cool down first to reach the instability region which starts at $T \simeq 10 \mbox{MeV}$. One cannot draw more conclusions because the CFL phase has several other contributions to be considered.

 The r-mode instability window for the CSL phase provides a wider stability region which ends at a temperature $T\sim 0.01 \mbox{MeV}$, see Fig~(\ref{fig:BV_nu_c_csl}). The situation is not fully resolved in the case of the CSL phase due to the uncertainties in the values of the energy gap and the critical temperature, and the fact that there are other dissipative phenomena to be considered.

The 2SC phase has a structure very similar to unpaired quark matter, see Fig.~(\ref{fig:BV_nu_c_2sc}), and the conclusions for the unpaired phase can be similarly drawn for the 2SC phase. This is because in the 2SC phase there is always an ungapped quark color. This means that $\frac 1 3$ of the quarks behave exactly as the unpaired quark matter phase. Moreover, the strange quark always remains gapless. This leads to a suppression for the non-leptonic weak interaction close to $\frac 2 9$, for the strange quark \emph{Urca} processes it is $\frac 2 3$, and for the down quark \emph{Urca} processes it is $\frac 1 3$. Ref.~\cite{Madsenprl2} suggests that the 2SC phase is marginally inconsistent with LMXB's data, we have shown that the 2SC is completely consistent with the  LMXB's observations, see Fig.~(\ref{fig:BV_nu_c_2sc}).

 To better understand the reasone for the enhancement of the bulk viscosity (and of the r-mode stability window) due to color superconductivity, we need to look deeper into the behavior of the bulk viscosity with respect to the oscillation frequency $\omega$ and the difference in interaction rates $\lambda_i$. It is hard to see that from Eq.~(\ref{zeta1-3}). Let us consider the case of a single-interaction. If only ruled by the non-leptonic weak interaction, the bulk viscosity takes the form \cite{SSR2}

\begin{equation}
 \zeta_{\rm non} \simeq \frac{\lambda_{1}C_{1}^2}{\omega^2+\left(\lambda_{1}A_{1}/n\right)^2}.
\end{equation}
 
 As can be seen from this formula, there are two regions with two different dependences on the interaction rate: 1) The region where $\omega^2  \gg \left(\lambda_{1}A_{1}/n\right)^2$, in this region the bulk viscosity can be approximated by $ \zeta_{1}\simeq \frac{\lambda_{1}}{\omega^2}C_{1}^2$ and any reduction to $\lambda_1$ leads to a similar reduction of the bulk viscosity ($e^{-\frac{2 \Delta_{CFL}}{T}}$ in the CFL phase). 2) The region where $\omega^2  \ll \left(\lambda_{1}A_{1}/n\right)^2$, in this region the bulk viscosity is approximated by $\zeta_{1} \simeq \frac{n C_1^2}{A_1^2 \lambda_1}$ and any reduction to the interaction rate ($e^{-\frac{2 \Delta_{CFL}}{T}}$ in the CFL phase) leads to an enhancement of the bulk viscosity ($e^{\frac{2 \Delta_{CFL}}{T}}$).

 Another way of describing the argument above is to say the bulk viscosity is a resonance effect between the density oscillation on one side, and the interaction rates that try to restore $\beta$-equilibrium during these oscillations on the other \cite{Alford:2006gy}.
 
 The thorough reader might notice that our results for the normal quark matter are a bit different -albeit close- to those presented in Ref.~\cite{Madsenprl2}, there are several reasons for this discrepancy, unlike Ref.~\cite{Madsenprl2}, which considers the bulk viscosity only due to the non-leptonic weak interaction, the bulk viscosity coefficient considered here takes into account the contribution of both the non-leptonic and the \emph{Urca} processes. Another source of discrepancy was due to the assumptions used to calculate the chemical potentials of quarks, in here we determine the chemical potential of each quark species from the charge neutrality and $\beta$-equilibrium conditions in a fashion similar to Ref.~\cite{SSR2}.

 This manuscript represents a first step towards a more complete treatment of the r-mode instabilities of color-superconducting quark matter. As a follow up to this work, one could attempt to calculate other dissipative phenomena in deconfined quark matter, such as the shear viscosity in superfluid phases like the CFL \cite{cristina_shear} and the CSL phase, the bulk viscosity due to superfluidity in the CFL \cite{cristina_bulk} and the CSL phase, see also Ref.~\cite{Mannarelli:2008jq}, or dissipation due to electron-electron scattering in the system, or due to surface friction between a rotating core and a nearly static crust in the case of hybrid stars. Supplementing these results with the ones in this article should provide us with more accurate figures for the r-mode instability window in the superfluid phases, which would then allow us to draw more firm astrophysical conclusions.

 In this work a constant density approximation of the star was assumed, however, one can, in principle, generalize the calculations to include different models that might result in a density profile of the star, which will lead to a radis-dependent viscosities and dissipative timescales. 

 While preparing this manuscript, an article which addresses the dissipation of r-mode instabilities in the CFL phase due to the viscosities calculated in Refs.~\cite{cristina_shear, mark_bulk} was published, Ref.~\cite{Jaikumar:2008kh}. Some discrepancies between their treatment and the one presented here need to be pointed out, first, in order to calculate the dissipative timescale associated with bulk viscosity they used the approximation that was used in Ref.~\cite{lindblom_owen_morsink,kokkotas_stergioulas} in which the Lagrangian perturbation of the fluid is approximated by an Eulerian perturbation. This approximation greatly overestimates the dissipative timescale of the bulk viscosity, as was shown in Refs.~\cite{Anderssonreview, Lind2nd}. Another point worth mentioning was that the shear viscosity formula that was used is different, it has a temperature dependence of $T^{-2}$ which was derived in \cite{Haensel:1989ja}, however, it was shown in \cite{Heiselberg:1993cr} that a more proper treatment leads to the formula we have used in here, \emph{i.e.} Eq.(\ref{sqm-sv}).

 \section*{Acknowledgements}

 The author acknowledges discussions with D.~Rischke, I.~Shovkovy, J.~Schaffner-Bielich, M.~Alford, J.~Noronha,
A.~Schmitt, C.~Manuel, M.~ Mannarelli, and J.~Madsen. B.A.S. also acknowledges support from the Frankfurt International Graduate School of Science (FIGSS) and the hospitality of the Institut de Ci\`encies de l'Espai (IEEC-CSIC). 
\bibliographystyle{apsrev}
\bibliography{rmode}
\end{document}